# Generation and propagation of topological solitons in a chain of coupled parametric-micromechanical-resonator arrays

Hiroshi Yamaguchi and Samer Houri

*NTT Basic Research Laboratories, NTT Corporation, Atsugi-shi, Kanagawa 243-0198, Japan*
(*hiroshi.yamaguchi.zc@hco.ntt.co.jp*)

Using a coupled parametric-resonator array for generating and propagating a topological soliton in its rotating-frame phase space is theoretically and numerically investigated. In an analogy with the well-known $\phi^4$ model, the existence of a soliton is topologically protected as the boundary of two different phase domains of parametric oscillation. Numerical simulation indicates that the propagation can be triggered by switching of the phase state of one specific resonator, and the effects of damping, collision, and the symmetry lifting by harmonic drive on the propagation dynamics are studied. The topological soliton can be implemented by using electromechanical resonators, which allow its propagation dynamics to be precisely electrically controlled and provide a fully controlled on-chip test bed for the study of a topological soliton.

I. Introduction

A parametric resonator is one of the simplest Floquet systems [1]-[3] that spontaneously breaks time translational symmetry [4]-[6]. Harmonic resonance forces a resonator to oscillate near the eigenfrequency, while parametric modulation at twice the eigenfrequency forces the system to maintain half-period time translational symmetry. Consequently, excited oscillations break the externally defined half-period symmetry and drop into one of the doubly degenerated phase states, each of which is transformed from the other through the half-period time translation. The parametric resonator is frequently described by a pseudo-energy Hamiltonian in its rotating frame, and the two oscillation states correspond to two local minima of a double-well-shaped potential [7][8]. It can be implemented by using various physical systems, ranging from a macroscopic child swing, optical nonlinear systems [9]-[11], and LCR circuits [12][13] to state-of-the-art microscopic on-chip devices like micro- and nano-mechanical systems [14]-[17] and superconducting circuits [18]-[19]. As for these systems, oscillation states can be externally controlled by artificially "manipulating" the double-well potential. Using the controllability, applications such as noise squeezing [14][18]-[20], computations [13][17][21], and a symmetry-lifting detector for quantum measurement [22]-[24] have been proposed and experimentally demonstrated.

Reflecting the recent progress of device-fabrication technology, interest in a "coupled mechanical-resonator array" has been increasing. Devices comprising a large number of micro- and nano-electromechanical systems are routinely fabricated, and peculiar collective behaviors, such as mode localization and synchronization, have been studied both theoretically and experimentally [25]

-[40] . Among those behaviors, intrinsic localized modes (ILMs), also known as lattice solitons, have been studied [26] ,[32] -[40] . A parametric-resonator array has also been extensively studied and well-described in the rotating frame by using the so-called "parametrically driven nonlinear Schrodinger equation" (PNSE) [32] -[40] , which was shown to have solitons as solutions. However, the study of topological solitons in a parametric-resonator array is limited, and, to the best of my knowledge, only the stability of a static dark soliton has been discussed [39] [40] . As in the case of other topological edge states, a topological soliton is a kind of topological defect between two different phase domains, and is a commonly observed excitation in one-dimensional systems with discontinuous symmetry breaking [41] . The existence of these solitons is robust in the sense that it is topologically protected against external perturbation, so it has the potential to be an important mode of robust device operation. Accordingly, in this paper, we study topological solitons in a one-dimensional array of parametric resonators. In particular, we numerically study the generation and following propagation dynamics of topological solitons, which can be directly demonstrated with real devices. Considering the results of this study, we discuss the feasibility of using a realistic coupled-electromechanical-resonator array to experimentally demonstrate the robustness of topological solitons.

II. Model

The model used for this study is a coupled one-dimensional parametric-resonator array. The equation of motion is given by

$$\ddot{q}_n = -\gamma \dot{q}_n - \omega_0^2[1 + 2\Gamma \cos(2\omega_0 t)]q_n - \alpha q_n^3 + g\omega_0(q_{n+1} - q_n) + g\omega_0(q_{n-1} - q_n).  \tag{1}$$

where $q_n$ is displacement of the *n*-th resonator (*n*: integer) from its equilibrium position, $\omega_0$ is resonant (angular) frequency, $\gamma$ is the damping coefficient, $\Gamma$ is parametric excitation amplitude, $\alpha$ is the Duffing nonlinear coefficient, and $g$ is the coupling constant between two adjacent resonators. For simplicity, no detuning is assumed, i.e., parametric excitation is at exactly twice the resonance frequency, and $\gamma$, $\omega_0$, $\Gamma$, $\alpha$, and $g$ are assumed to be identical for all resonators. The implemented model based on an electromechanical parametric-resonator array is shown schematically in Fig. 1. Suspended parametric resonators are coupled through the overhang formed through the fabrication process [25] [26] [42] [43] . Parametric oscillation is electrically excited by applying an alternate voltage to the common electrode. A strong coupling regime, i.e. $\gamma < g$, is assumed, and rotating-frame approximation is applied by introducing slowly varying quadrature amplitudes, $c_n(t)$ and $s_n(t)$, defined by

$$q_n(t) = Re[(c_n(t) + is_n(t))e^{i\omega_0 t}] = c_n(t)\cos\omega_0 t - s_n(t)\sin\omega_0 t. \qquad (2)$$

The equations of motion for new variables become

$$\begin{aligned}
\frac{dc_n}{dt} &= -\frac{\gamma}{2}c_n + \frac{\omega_0 \Gamma}{2}s_n - \frac{3\alpha}{8\omega_0}s_n(c_n^2 + s_n^2) + \frac{g}{2}(s_{n+1} - 2s_n + s_{n-1}), \\
\frac{ds_n}{dt} &= -\frac{\gamma}{2}s_n + \frac{\omega_0 \Gamma}{2}c_n + \frac{3\alpha}{8\omega_0}c_n(c_n^2 + s_n^2) - \frac{g}{2}(c_{n+1} - 2c_n + c_{n-1}).
\end{aligned} \qquad (3)$$

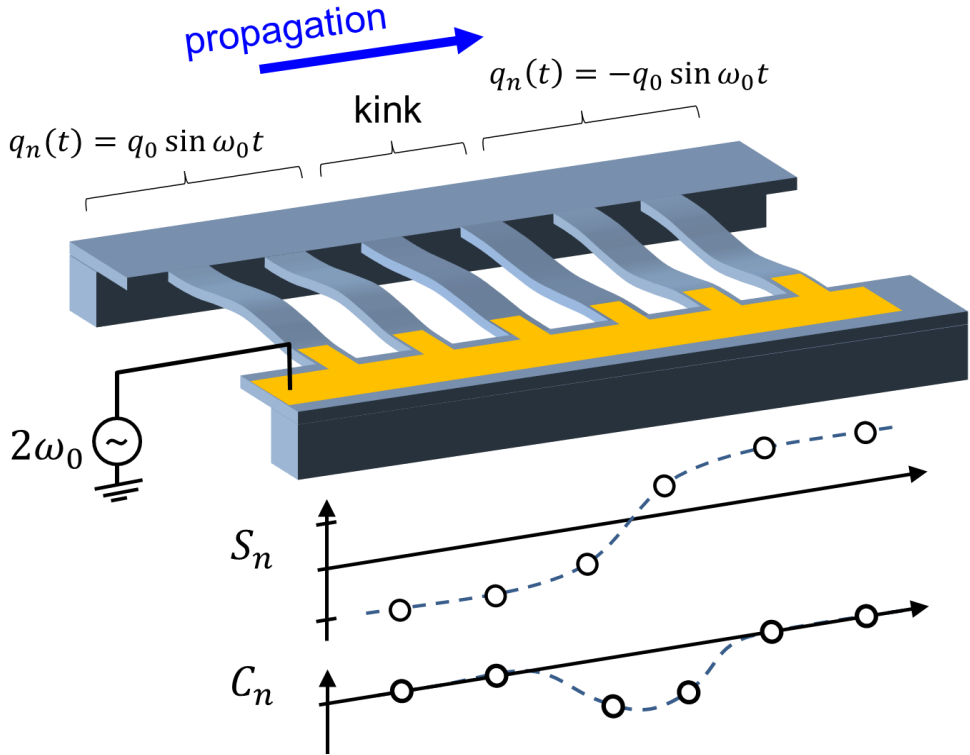

Fig. 1
Schematic drawing of a coupled electromechanical parametric-resonator array and a travelling kink (topological soliton) as a boundary between two different oscillation phase domains. Parametric oscillations are simultaneously excited by applying a voltage to the common electrode (yellow) at twice the resonance frequency, $2\omega_0$. Each resonator is coupled with two adjacent resonators through an overhang formed by selective etching [25] [26] [42] [43]. Also shown are two quadrature amplitudes, $S_n \equiv s_n/\rho_0$ and $C_n \equiv c_n/\rho_0$, for a travelling kink solution. Cosine component $C_n$ has maximum amplitude at the kink position, which is proportional to the traveling velocity of the soliton.

The dynamics in the absence of damping ($\gamma \sim 0$) can be derived from the canonical equations of

motion as

$$\frac{ds_n}{dt} = \frac{\partial H_r}{\partial c_n}, \quad \frac{dc_n}{dt} = -\frac{\partial H_r}{\partial s_n}, \quad (4)$$

where $H_r$ is the rotating-frame Hamiltonian, given as

$$\begin{aligned} H_r(s_n, c_n) &= \sum_n H_n + H_{int} \\ &\equiv \sum_n \left[ -\frac{\omega_0 \Gamma}{4}(s_n^2 - c_n^2) + \frac{3\alpha}{32\omega_0}(s_n^2 + c_n^2)^2 \right] \\ &\quad + \frac{g}{4} \sum_n [(s_{n+1} - s_n)^2 + (c_{n+1} - c_n)^2]. \end{aligned} \quad (5)$$

A three-dimensional plot of $H_n$ for one specific resonator $n$ is shown in Fig. 2, indicating the features of a double-well potential [8]. The two local minima $(s_n, c_n) = (\pm \rho_0, 0)$, with $\rho_0 = 2\omega_0 \sqrt{\frac{\Gamma}{3\alpha}}$, correspond to two steady-state phase states, $q_n(t) = \mp \rho_0 \sin \omega_0 t$. The parametrically amplified oscillation amplitude $s_n$ saturates at the value of $\pm \rho_0$ according to Duffing nonlinearity.

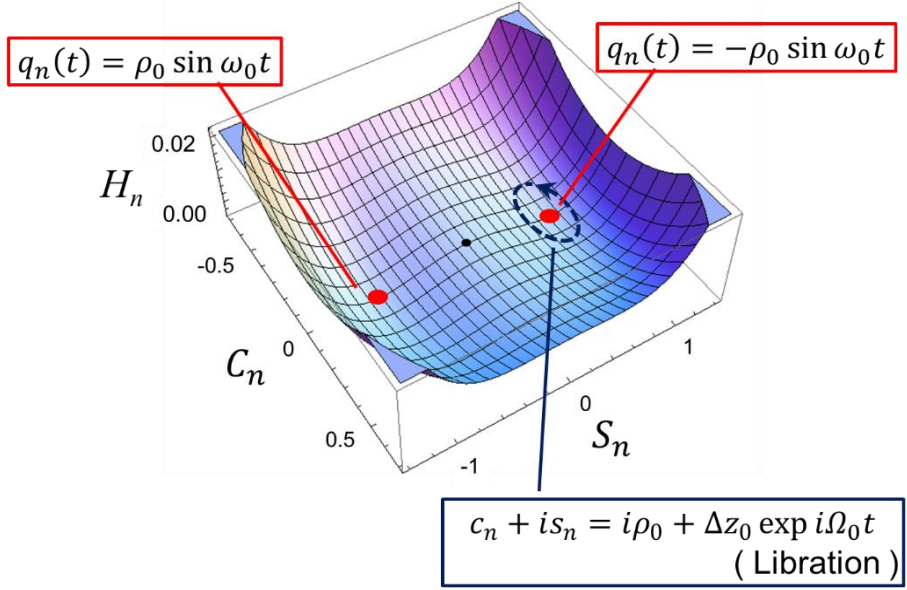

Fig. 2
Three-dimensional plot of rotating-frame Hamiltonian $H_n$, calculated with $\alpha = 0.02$, $\omega_0 = 1$, $\Gamma = 0.01$. The two fixed points corresponding to parametric-oscillation sates are shown by red markers at $(s_n, c_n) = (\pm \rho_0, 0)$ with $\rho_0 = 2\omega_0 \sqrt{\Gamma/3\alpha} = 0.82$. The dashed closed orbit shows the trajectory of libration.

To intuitively find the effect of coupling, the continuous variables $(s_n, c_n)$ are approximately replaced by discrete pseudo spin variables, $\sigma_n \equiv s_n/\rho_0$, which have only two discrete values, $\sigma_n = \pm 1$, to obtain the effective Hamiltonian $H_{eff}$ for $\sigma_n$ as

$$H_{eff}(\sigma_n) \equiv H_r(\rho_0 \sigma_n, 0) = -\frac{g\rho_0^2}{2}\sum_n \sigma_n \sigma_{n+1} + Cst. \quad (\sigma_n = \pm 1), \tag{6}$$

where $Cst.$ is a $\sigma_n$-independent constant. Equation (6) is identical to a 1D Ising Hamiltonian [21] [43]. The analogy suggests that the phase boundary can exist at any finite temperature because the 1D Ising Hamiltonian does not show phase transition. In Section 4, it is more accurately shown that (i) the phase boundary of parametric-resonator oscillation states corresponds to a topological soliton in the phase space and (ii) the dynamics can be controlled by externally modifying the parameters of parametric excitation.

Before the soliton solution is discussed, the weak linear excitation around the synchronized parametric oscillation is first described and used to find the stability and fundamental properties of coupled-parametric-resonator arrays. Without coupling $g$, each resonator amplitude $(s_n, c_n)$ sits at one of two steady-state fixed points $(\pm \rho_0, 0)$ in the phase space (red makers in Fig. 2). When the resonator is forced to oscillate by an external perturbation, it departs from the fixed point to start a slow periodic oscillation around that point (dashed line) [44]. This rotating-frame orbit is referred to as "libration" [34], in an analogy with slow periodic astronomic motions. It is then considered that all parametrically oscillating resonators are synchronized through the coupling to the positive phase state, i.e., $(s_n, c_n) = (+\rho_0, 0)$ for all values of $n$, and small deviation $\Delta z_n$ is excited in its vicinity, i.e., $\Delta z_n \equiv c_n + i(s_n - \rho_0)$. It can be seen that the libration around the fixed point is also correlated and propagates as a traveling wave through the coupling.

By linearizing Equation (3), the equation of motion for complex variable $\Delta z_n$ becomes

$$\frac{d\Delta z_n}{dt} = i\omega_0 \Gamma \Delta z_n - i\frac{g}{2}(\Delta z_{n+1} - 2\Delta z_n + \Delta z_{n-1}). \tag{7}$$

which is the wave equation for a coupled "librator" array with on-site frequency of $\Omega_0 = \omega_0 \Gamma$, and coupling $g$. The libration waves (LWs) are stable and propagate along the resonator array if all the resonators are synchronized in one specific oscillation phase. Under the assumption of a travelling-wave solution, namely, $\Delta z_n \sim \exp i(\Omega(k)t - nk\Delta x)$, where $\Delta x$ is adjacent resonator distance and $k$ is wave number, the dispersion relation is given by

$$\Omega(k) = \Omega_0 - g(\cos k\Delta x - 1). \tag{8}$$

Group velocity can be obtained as

$$v_g = \frac{\partial \Omega}{\partial k} = \Delta x g \sin k\Delta x. \tag{9}$$

It therefore follows that $0 \leq |v_g| \leq \Delta x g \equiv v_{g,max}$. In section IV, our numerical analysis confirms that as well as the soliton, a wave packet of travelling LWs, with velocity upper bound $v_{g,max}$, is also generated upon application of an external excitation, i.e. the phase state switching in one specific resonator.

III. Continuous model and analytical solutions

When considered along the s-axis, the double-well potential shown in Fig. 2 is similar to that of $\phi^4$ model [41]. The Hamiltonian of $\phi^4$ model is given by

$$H_{\phi^4} = \int \left[\frac{1}{2}\pi^2 + \left(\frac{\partial \phi}{\partial x}\right)^2 + V(\phi)\right] dx, \qquad V(\phi) = \frac{1}{4}(\phi^2 - 1)^2. \tag{10}$$

where $\phi(x,t)$ is a field variable, and $\pi(x,t) = \partial \phi(x,t)/\partial t$ is its canonically conjugate momentum. Potential energy $V(\phi)$ is double-well-shaped with the minima at $\phi = \pm 1$, and it is well known that the model has a topological soliton solution. If the boundary condition, $\lim_{x \to \pm \infty} \phi(x) = \pm 1$, is applied, a phase boundary, called a "kink" or "topological defect," which can propagate at arbitral velocity, must exist between the positive and negative domains. Strictly speaking, the phase boundary is not a soliton, and it is often referred to as "quasi-soliton," because of the lack of waveform conservation when two excitations collide [41]. This is because upon the collision the energy is transferred to internal mode excitations. Hereafter, it is shown that a similar topological (quasi-) soliton exists as a phase boundary between two different oscillation states in a parametric-resonator array. The significant difference between the parametric-resonator array and the $\phi^4$ model is that the phase boundary is in rotating-frame quadrature amplitudes in the former, whereas is in laboratory-frame coordinates in the latter. Therefore, while oscillation amplitude is finite over nearly the whole range of $x$, the phase of the oscillation has a soliton-like behavior in parametric-resonator array. This type of soliton is therefore often called a "dark" or "grey" soliton, which plays a role also in nonlinear fiber optics [45].

A continuous model is used here to describe a static soliton solution. The Hamiltonian can be obtained from Equation (5) by replacing the variables as follows:

$$s_n \to s(x)\sqrt{\Delta x}, \qquad c_n \to c(x)\sqrt{\Delta x},$$
$$s_{n+1} - s_n \to \Delta x \frac{\partial s(x)}{\partial x}, \qquad c_{n+1} - c_n \to \Delta x \frac{\partial c(x)}{\partial x}$$

$$\sum_n X_n \to \int \frac{dx}{\Delta x} X(x).$$

Hereafter, to simplify the equation, dimensionless time and length units, i.e., $\omega_0 = 1$ and $\Delta x = 1$, are used. It thus follows that

$$H_r(s,c) = \int dx \left[ -\frac{\Gamma}{4}(s^2 - c^2) + \frac{3\alpha}{32}(s^2 + c^2)^2 \right] + \frac{g}{4} \int dx \left[ \left(\frac{\partial s}{\partial x}\right)^2 + \left(\frac{\partial c}{\partial x}\right)^2 \right], \tag{11}$$

and the canonical field equations become

$$\begin{aligned}\frac{\partial s}{\partial t} &= \frac{\delta H_r(s,c)}{\delta c} = \frac{\Gamma}{2}c + \frac{3\alpha}{8}(s^2 + c^2)c - \frac{g}{2}\left(\frac{\partial^2 c}{\partial x^2}\right), \\ \frac{\partial c}{\partial t} &= -\frac{\delta H_r(s,c)}{\delta s} = \frac{\Gamma}{2}s - \frac{3\alpha}{8}(s^2 + c^2)s + \frac{g}{2}\left(\frac{\partial^2 s}{\partial x^2}\right).\end{aligned} \tag{12}$$

By introducing complex variables, $z(x,t) \equiv c(x,t) + is(x,t)$, Equation (12) becomes a parametrically driven Nonlinear Schrodinger equation (PNSE) [32]-[40]

$$i\frac{\partial z}{\partial t} = \frac{g}{2}\frac{\partial^2 z}{\partial x^2} - \frac{3\alpha}{8}|z|^2 z - \frac{\Gamma}{2}\bar{z}. \tag{13}$$

The stability of bright and dark soliton solutions as the ILMs for several parameter ranges has been discussed [39] [40]. Zero-frequency dark mode solitons are focused on hereafter. The boundary condition of dark mode is given by

$$\lim_{x \to \pm\infty} s(x,t) = \pm\rho_0, \quad \lim_{x \to \pm\infty} c(x,t) = 0. \tag{14}$$

The time-independent kink solution satisfying the boundary condition is given as [36]

$$s(x,t) = \pm s_{kink}(x) \equiv \pm\rho_0 \tanh\left(\frac{x - x_0}{\lambda}\right), \quad c(x,t) = 0. \tag{15}$$

Solutions with positive and negative signs in Equation (15) are respectively referred to as a "kink" and "antikink." They correspond to the phase boundary between two parametric-oscillation states (see Fig. 1) and a continuous version of a dark soliton obtained in [39] and [40] for small coupling limit. Kink length $\lambda$ is given by

$$\lambda = \sqrt{\frac{2g}{\Gamma}}. \tag{16}$$

Equation (16) means that stronger coupling gives a longer kink. This finding is intuitively reasonable because stronger coupling makes the effect of the neighboring resonator stronger, leading to the wider phase-transition region. From Eq. (11), pseudo-energy density $\varepsilon_r$ is given by

$$\begin{aligned}\varepsilon_r &= -\frac{\Gamma}{4}(s^2 - c^2) + \frac{3\alpha}{32}(s^2 + c^2)^2 + \frac{g}{4}\left[\left(\frac{\partial s}{\partial x}\right)^2 + \left(\frac{\partial c}{\partial x}\right)^2\right] \\ &= \frac{\Gamma^2}{6\alpha}\left[-2\tanh^2\left(\frac{x-x_0}{\lambda}\right) + \tanh^4\left(\frac{x-x_0}{\lambda}\right) + \frac{1}{2\cosh^4\frac{x-x_0}{\lambda}}\right].\end{aligned} \tag{17}$$

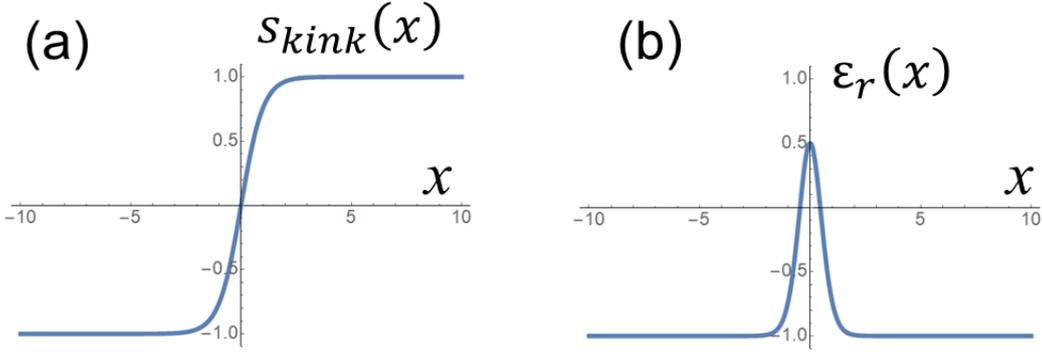

**Fig. 3**
(a) Waveform $s_{kink}(x)$ and (b) pseudo-energy density $\varepsilon_r(x)$ of the static kink solution as a function of normalized position $x$, where the units used are $\rho_0 = 1$, $\lambda = 1$, and $\Gamma^2/6\alpha = 1$.

The waveform and pseudo-energy density of the kink are plotted in Fig. 3(a) and (b). The additional energy is located around the kink position. Due to the space-translational symmetry, the equation has solution of zero-frequency excitation, corresponding to a Nambu-Goldstone mode given by $\frac{\partial s_{kink}(x)}{\partial x} \sim \text{sech}^2\left(\frac{x-x_0}{\lambda}\right)$; i.e.,

$$s(x,t) = s_{kink}(x) + \kappa \,\text{sech}^2\left(\frac{x-x_0}{\lambda}\right), \qquad c(x,t) = 0 \tag{18}$$

satisfies Equation (12) within the linear approximation in $\kappa$.

In contrast to the $\phi^4$ model, Equation (12) is not relativistically invariant. Therefore, an analytical solution with finite travelling velocity cannot be obtained by Lorentz transformation to a moving frame, so the solution is numerically calculated instead. Under the assumption of a travelling wave form as

$$s(x,t) = \rho_0 S(z), \qquad c(x,t) = \rho_0 C(z), \qquad z = x - vt, \tag{19}$$

Equation (12) becomes

$$\eta S''(z) + \mu C'(z) + S(z) - S(z)^3 - S(z)C(z)^2 = 0 \\ \eta C''(z) - \mu S'(z) - C(z) - C(z)^3 - C(z)S(z)^2 = 0. \tag{20}$$

where

$$\eta = \frac{g}{\Gamma} = \frac{\lambda^2}{2}, \qquad \mu = \frac{2v}{\Gamma}. \tag{21}$$

The boundary condition given by Equation (14) becomes

$$\lim_{z \to \pm\infty} S(z) = \pm 1, \lim_{z \to \pm\infty} C(z) = 0. \tag{22}$$

The solution for $\eta = 1$ is then obtained because it is straightforward to obtain a general solution by rescaling, i.e., $z \to z/\sqrt{\eta}$ and $\mu \to \mu/\sqrt{\eta}$. Equation **(20)** thus becomes

$$S''(z) + \mu C'(z) + S(z) - S(z)^3 - S(z)C(z)^2 = 0, \\ C''(z) - \mu S'(z) - C(z) - C(z)^3 - C(z)S(z)^2 = 0. \tag{23}$$

If kink velocity $\mu$ is small enough, the first-order perturbation approximation, given as

$$S(z) = \tanh\frac{z}{\sqrt{2}} + \mu S_1(z) + O(\mu^2), \qquad C(z) = \mu C_1(z) + O(\mu^2), \tag{24}$$

can be used with the boundary condition, namely,

$$\lim_{z\to\pm\infty} S_1(z) = 0, \lim_{z\to\pm\infty} C_1(z) = 0. \tag{25}$$

Equation (23) thus becomes

$$S_1''(z) + \left(1 - 3\tanh^2 \frac{z}{\sqrt{2}}\right) S_1(z) = 0,$$

$$C_1''(z) - C_1(z) = \tanh^2 \frac{z}{\sqrt{2}} C_1(z) + \sqrt{\frac{1}{2}} \operatorname{sech}^2 \frac{z}{\sqrt{2}}. \tag{26}$$

The upper equation in (26) has solution $S_1(z) = \kappa \operatorname{sech}^2 \frac{z}{\sqrt{2}}$ with arbitral coefficient $\kappa$. This solution corresponds to the Nambu-Goldstone mode [Equation (18)] and simply provides a spatially displaced solution, so $S_1(z) = 0$ can be set without loss of generality. The lower equation can be numerically solved, and the solution thereby obtained is shown in Fig. 4(a). The travelling kink solution has non-zero $C$-quadrature, and its amplitude is proportional to normalized velocity $\mu$. The existence of the travelling-wave solution indicates that the waveform of the kink is preserved as the kink propagates. The kink and antikink can therefore be regarded as topological solitons.

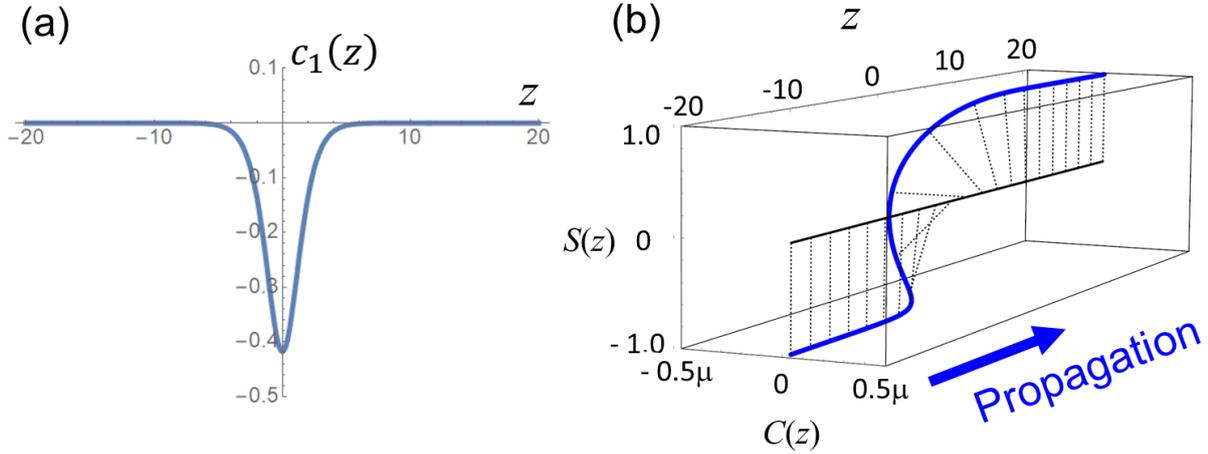

**Fig. 4**
(a) Numerically calculated $C_1(z)$. (b) Three-dimensional plot of $S(z)$ and $C(z)$ for a travelling-kink solution (blue solid line). The dashed lines are guides for the eye. The solution has a clockwise rotation in phase space along the propagation axis.

A three-dimensional plot of the calculated waveform is shown in Fig. 4(b). The solution has a clockwise rotation in phase space along the propagation axis. In contrast to the other topological solitons, where the phase rotation around the kink position is in the laboratory frame, the rotation is in the rotating-frame phase space. Because the oscillation phase at positive and negative infinities is constrained to zero and $\pi$, respectively, the boundary cannot disappear. This property protecting the existence of the kink plays a similar role as topological edge states. Number of rotations in phase space can thus be similarly defined as a topological charge, namely,

$$Q_\theta \equiv \frac{1}{2\pi} \int_{z=-\infty}^{+\infty} \left(S\frac{dC}{dz} - C\frac{dS}{dz}\right)\frac{dz}{r^2}, \qquad (27)$$

which has a value of $\text{sgn}(\mu)/2$ for a kink solution. For an antikink solution, both $S(z)$ and $C(z)$ reverse signs, so the topological charge has the same sign as a kink. If the helicity, i.e., the number of windings in the propagation direction, is defined as

$$h \equiv \frac{Q_\theta}{\text{sgn}(\mu)}, \qquad (28)$$

$h$ is always a positive half. This result simply reflects the fact that the rotation direction around a local-minimum fixed point is always anticlockwise in rotating-frame phase space.

IV. Discrete model and numerical analysis

To study the detailed propagation dynamics of topological solitons, the time evolution of two quadrature amplitudes was numerically calculated by using a discrete model. The stability of a static solution with no damping limit was studies by similar calculations [39] and [40] . Hereafter, to discuss the experimental feasibility of using a micro-electromechanical resonator array, a travelling wave with finite damping is studied in detail. This study focuses on the following three points of interests. The first point is the effect of damping on the soliton propagation. To experimentally demonstrate the soliton generation and study the propagation dynamics, the soliton should propagate long enough to be observed in a real resonator array with given damping. The second point is a scheme to excite the soliton. In real experiments, switching of an oscillating phase can be induced by electromechanical transduction means [23] [46] . It is confirmed hereafter that a travelling soliton can be generated by phase switching and can be driven by additional harmonic excitation. Such driving is also important to initialize all phase states. The third point is the collision dynamics of two solitons. This point is important because one of the required properties of a soliton is conservation of wave form after the collision process.

As for the numerical study, the discrete model given by Eq. (3) is used, and the time evolution of two quadrature amplitudes is calculated by using a standard Runge–Kutta method**.** As in the case of a continuous model, dimensionless amplitudes are introduced as

$$s_n = \rho_0 S_n, \qquad c_n = \rho_0 C_n. \tag{29}$$

A set of equations to be numerically solved is then obtained as

$$\frac{dC_n}{dt} = -\frac{\gamma}{2}C_n + \frac{\Gamma}{2}S_n[1 - C_n^2 - S_n^2] + \frac{g}{2}[S_{n+1} - 2S_n + S_{n-1}],$$
$$\frac{dS_n}{dt} = -\frac{\gamma}{2}S_n + \frac{\Gamma}{2}C_n[1 + C_n^2 + S_n^2] - \frac{g}{2}[C_{n+1} - 2C_n + C_{n-1}]. \tag{30}$$

Stability of the static and propagating solitons is first confirmed. Time evolutions of $S_n$ and $C_n$ when $g = 0.02$ and $\Gamma = 0.01$ are calculated using the initial conditions,

$$S_n(t=0) = \tanh\left(\frac{n}{\lambda}\right), \qquad C_n(t=0) = -0.42\mu \operatorname{sech}^2 \frac{n}{1.2\lambda}. \tag{31}$$

These two initial waveforms are approximately obtained from continuous model analysis, i.e., Eq. (15), and the hyperbolic fitting curve applied to the numerical solution given in Fig. 4(a), respectively. The results of the calculation are shown in Fig. 5, which indicates that the kink propagates to the positive/negative direction when given $\mu$ is positive/negative, as expected from the continuous model analysis in the previous section. It confirms that the propagation velocity $v_{kink}$ calculated from the numerically obtained $S_n(t)$ and $C_n(t)$ is proportional to parameter $\mu$, as $v_{kink} \sim 0.0069\mu$, as expected from the continuous-model analysis.

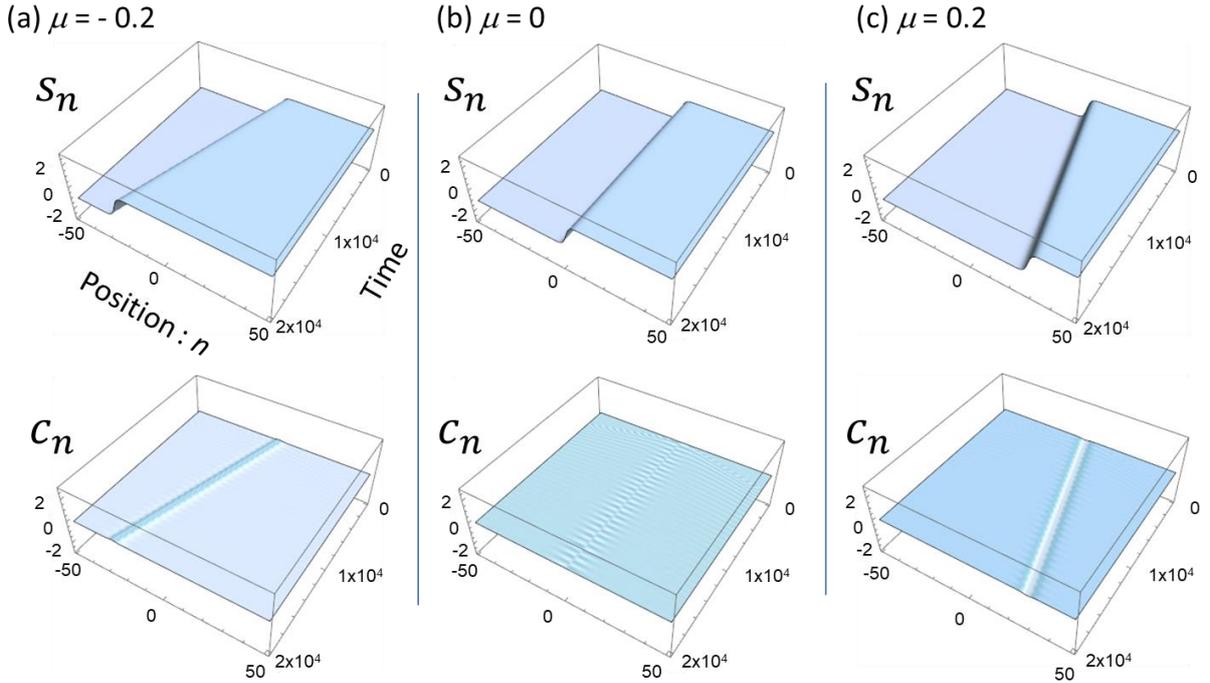

**Fig. 5**
Time evolutions of $S_n$ and $C_n$ calculated by using $g = 0.02$, $\Gamma = 0.01$, without damping, for (a) $\mu = -0.2$, (b) $\mu = 0$, and (c) $\mu = 0.2$, under the initial condition given by Equation (15) and the fitting curve of Fig. 4(a).

Next, the time evolution when the phase of one single resonator is switched is calculated as follows. The calculation starts from the initial condition as $S_n = -1, C_n = 0$ for all values of $n$. Polarity of $S_0$ is then reversed to +1 at $t = 0$, and the subsequent time evolution in $C_n$ and $S_n$ is calculated. This scenario corresponds to a feasible experiment, where the phase states of one parametric resonator are switched by an external drive signal [23] [46] . The results when $g = 0.02$ and $\Gamma = 0.01$ with a finite damping $\gamma$ are shown in Fig. 6. The calculation results for different values of quality factor $Q \equiv 1/\gamma$ are compared to find the effect of damping.

First, it is confirmed that a kink in $S_n$, i.e. the topological soliton, is generated by switching phase states at $n = 0$ resonator, and it propagates in the positive direction while keeping its shape. In addition, amplitude $C_n$ has a peak at the kink position as expected from the continuous-model analysis. The straight lines indicated by blue arrows in Fig. 6 shows the generation of weak excitations, which are propagating at constant velocity. This wave packet corresponds to travelling LW discussed in the previous section and is also generated by phase switching. The wave packet has the velocity that equals maximum group velocity $v_{g,max}$. This is because the oscillation phase of only one resonator was reversed so that the shortest-wavelength component of LW was generated. The kink is initially generated at the $n = 0$ resonator and propagates with a slightly lower velocity than $v_{g,max}$, but it decelerates as it propagates. With increasing damping, the reduction of velocity becomes more significant, indicating that the two are correlated.

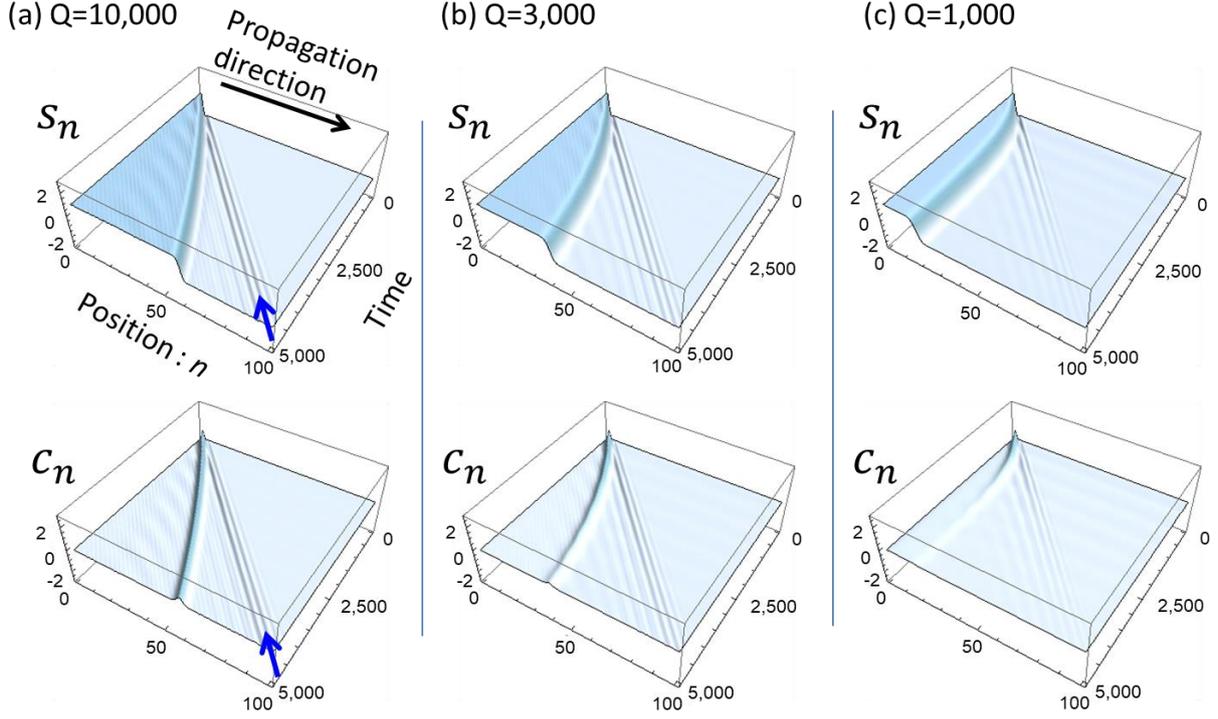

**Fig. 6**

Time evolutions of $S_n$ and $C_n$ calculated by using $g = 0.02$ and $\Gamma = 0.01$, and three different values of $Q$. Initially, all the resonators are set at the oscillation state $S_n = -1, C_n = 0$. At $t = 0$, variable $S_0$ was switched to $+1$ and the subsequent time evolution was calculated. Blue arrows show the wave front of the LW packet.

Calculated kink position $n_{kink}(t)$ is shown in Fig. 7(a) as a function of time $t$. The kink position is fitted by an exponential function of time, $n_{kink}(t) \sim a - be^{-t/\tau_r}$, and the obtained relaxation time $\tau_r$ is shown in (b). Here, $\tau_r$ is nearly equal to the inverse of damping, $Q$. Because peak amplitude of $C_n$ is proportional to velocity $\mu$ (see Eq. (24)), the damping in $C_n$ leads to deceleration of the soliton. The effect of damping is similar to that seen for sine-Gordon and $\phi^4$ models [41] . In a topological soliton, the existence of a kink is guaranteed by the topological property, so the kinetic part of its energy should be responsible for the energy dissipation.

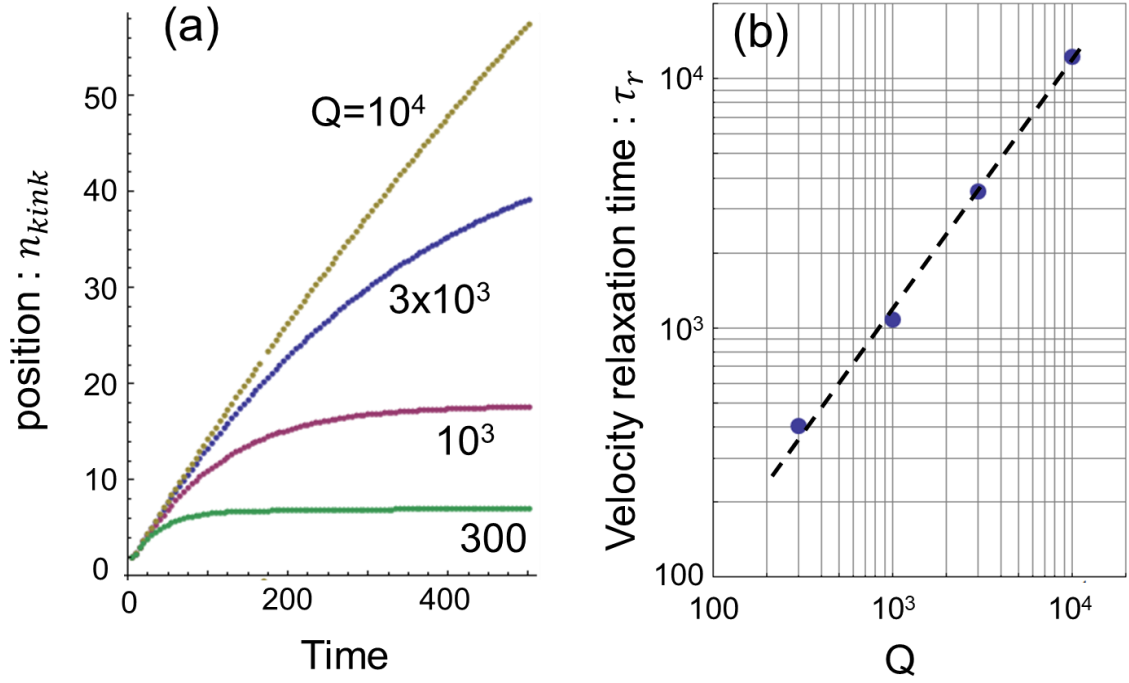

**Fig. 7**
(a) Calculated time dependence of kink position for four different values of $Q$. The kink exponentially decelerates. (b) Relaxation time of kink velocity $\tau_r$ numerically obtained from (a) as a function of $Q$. Relaxation time $\tau_r$ is nearly equal to the quality factor. The unit-slope dashed line is a guide for the eyes.

Hereafter, it is shown that propagation of the topological soliton can be driven by an external harmonic excitation. For a single parametric resonator, an additional harmonic drive breaks the discrete time translational symmetry with period $\pi/\omega_0$ and lifts the symmetry between two phase states [22] [23] [24] [46] . This symmetry lifting leads to the preference for one specific phase state, and the injection of noise activate the transition. In the case of a topological soliton, the symmetry lifting activates the transition from the unpreferred to preferred phase states around the kink position without the help of noise. The subsequent transition drives the propagation of the topological soliton.

For this study, we use the equation of motion with an additional harmonic drive, namely,

$$\ddot{q}_n = -\gamma \dot{q}_n - \omega_0^2[1 + 2\Gamma \cos(2\omega_0 t)]q_n - \alpha q_n^3 + g\omega_0(q_{n+1} - q_n) \\ + g\omega_0(q_{n-1} - q_n) + f_0 \sin \omega_0 t. \quad (32)$$

where $f_0$ is harmonic drive amplitude. Equation (30) thus becomes

$$\frac{dC_n}{dt} = -\frac{\gamma}{2}C_n + \frac{\Gamma}{2}S_n[1 - C_n{}^2 - S_n{}^2] + \frac{g}{2}[S_{n+1} - 2S_n + S_{n-1}] - F_0,$$
$$\frac{dS_n}{dt} = -\frac{\gamma}{2}S_n + \frac{\Gamma}{2}C_n[1 + C_n{}^2 + S_n{}^2] - \frac{g}{2}[C_{n+1} - 2C_n + C_{n-1}].$$
(33)

where $F_0 = f_0/4\rho_0$ is dimensionless harmonic-drive amplitude. The harmonic drive increases propagation velocity of the soliton until it saturates by damping. Equation (**33**) can be numerically solved as shown in Fig. 8. Applying the harmonic drive moves the kink position, whose propagation velocity saturates after damping time of $\sim Q$. As shown in Fig. 8(e), the saturation velocity is proportional to the harmonic drive amplitude and the quality factor. These results indicate that the topological soliton can be driven at constant propagation velocity under finite damping by simultaneously applying a harmonic drive to all the resonators. This symmetry lifting is practically feasible [23] [46] by applying an AC voltage at resonant frequency $\omega_0$ to the same electrode as that to which the parametric drive voltage was applied (Fig. 1). The continuous drive can sweep out all the kinks from the focused area and is useful for initializing the oscillation states to one specific phase.

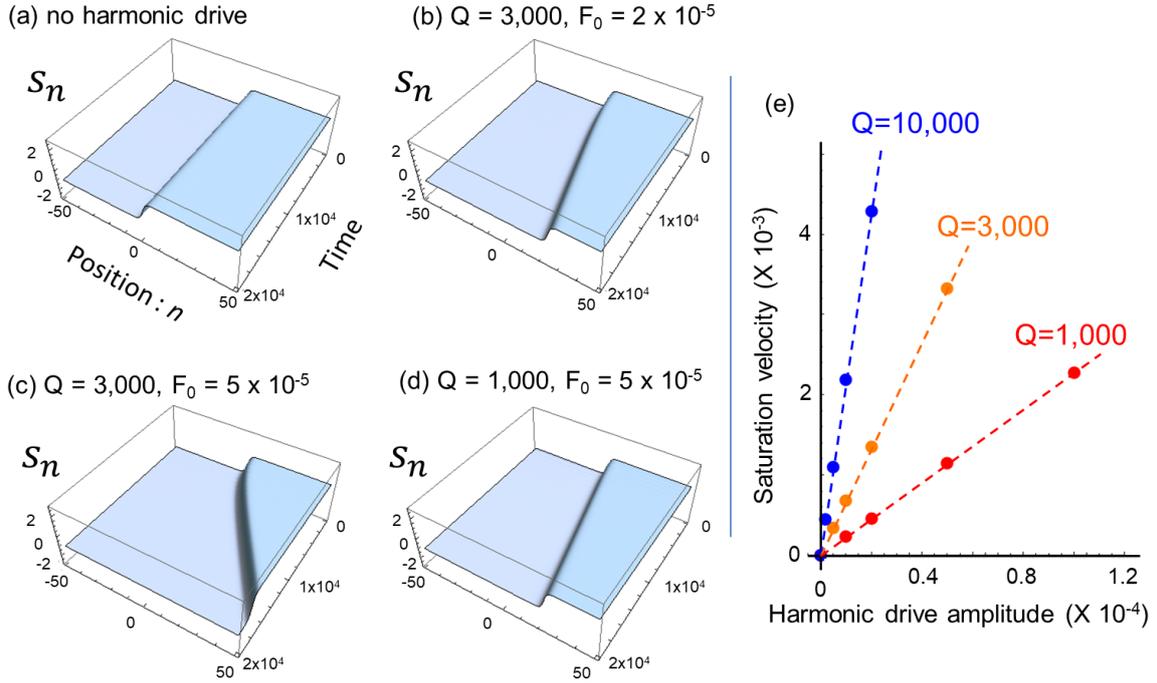

**Fig. 8**
(a)-(d) Calculated time evolutions of $S_n$ with a harmonic drive. Both the harmonic drive amplitude ($F_0$) and quality factor ($Q$) are varied. (e) Saturated propagation velocity as a function of harmonic drive amplitude ($F_0$). The same parameters as those in Fig. 5(b) were used.

Collision of two solitons is also numerically studied using the same model (Fig. 9). Two topological solitons, a kink and an anti-kink, are generated by phase switching at $t = 0$ and at $n = 0$ and $n = 100$ resonators. They then propagate in opposite directions and collide at $t \sim 5,000$. When damping is small enough (Fig. 9(a)), the two waves maintain enough energy at the collision to survive afterwards. However, if damping is increased (Fig. 9(b)) their energy is reduced at the collision, and the two solitons disappear after the collision. In their place, a strong ILM, or a "breather" mode [41], at $n \sim 50$ is generated by absorbing the energy associated with the solitons followed by outward propagation of LWs.

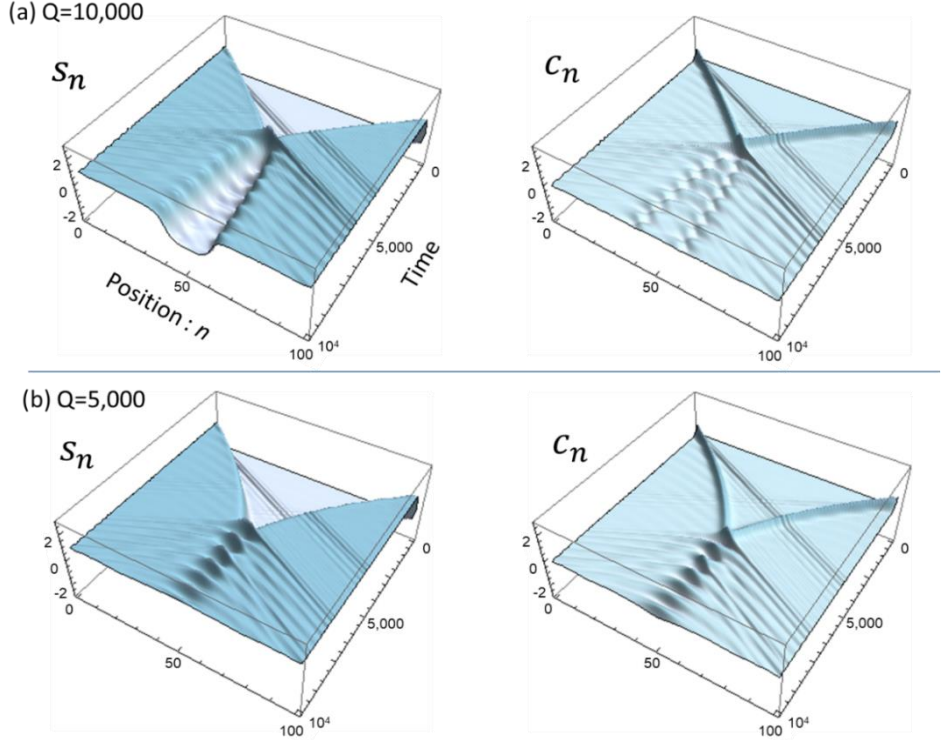

**Fig. 9**
Calculated time evolutions of $S_n$ and $C_n$ in the collision process of two solitons. Initially, all the resonators are set at oscillation state $S_n = -1, C_n = 0$. At $t = 0$, variables $S_0$ and $S_{100}$ were both switched to $+1$, and their subsequent time evolution was calculated. The same parameters as those for Fig. 6 were used.

Generation of a breather mode and LWs was observed also for $Q = 10,000$ (Fig. 9(a)), where each soliton velocity is reduced by the collision. Some portion of the soliton energy is transferred to the local mode and LWs, leading to reduction of velocity of the LWs. These results indicate that the waveform is not maintained after the collision, as in the case of the $\phi^4$ model, where the internal mode absorbs the energy to annihilate the two solitons. The topological soliton in this parametric-resonator array is therefore, rigorously speaking, not a soliton, but it can be referred to as a "quasi-soliton" as in the case of the $\phi^4$ model [41].

V. Discussion

Although the topological solitons discussed in this paper are not pure solitons, the existence of an isolated soliton is topologically protected, and they show experimentally interesting features. Especially, the solitons are formed in rotating frames, which can be electrically well controlled in practical devices. Based on the similarity between statically bistable structures, such as a buckled beam [47] and dynamic bistability in parametric resonators, similar topological solitons can be formed by using an array of coupled buckled beams. Such a system would be more accurately analogous to the $\phi^4$ model. However, because controlling static bistability is much more difficult than controlling dynamic bistability in parametric oscillators in real devices, our dynamical scheme allows more precise control and is promising for demonstrating on-chip test beds for propagation of topological solitons.

Two examples using nanoelectromechanical devices for experimentally demonstrating topological solitons are given hereafter. Our numerical calculation suggests that the typical parameters required to observe the soliton propagation are $Q > 10^3$, $g > 10^{-2}$, and $\Gamma \sim 10^{-2}$ for number of arrays of $\sim 100$. Here, $g$ can be estimated from frequency splitting between symmetric and antisymmetric vibration modes, and $\Gamma$ can be estimated from the depth of frequency modulation $\Delta\omega_0$ as $\Gamma \sim \Delta\omega_0/\omega_0$.

The first feasible example is given by the coupled doubly clamped beam array, already schematically shown in Fig. 1. In previous reports on two coupled parametric resonators [42][48], adjacent beams were coupled through an overhang formed by selective etching. Q is in the order of $10^3$ at room temperature and further increases up to $10^5$ by reducing the temperature. Frequency splitting between symmetric and antisymmetric modes is $\Delta f_{A-S} \sim 10$ $kHz$ as compared to resonance frequency $f_0 \sim 300$ kHz, so the estimated coupling is $g/\omega_0 = \Delta f_{A-S}/f_0 \sim 0.03$, which satisfies the requirement. However, the modulation depth of present device is in the order of a few kHz, therefore $\Gamma \sim 10^{-3}$. The numerical simulation with this parametric modulation depth shows that large-amplitude LWs are generated by phase switching, so it is difficult to clearly experimentally observe the generation of a soliton. Frequency-modulation depth should be increased by one order of magnitude by redesigning frequency-tuning efficiency. Alternatively, from the scale invariance of Eq. (3) under $\gamma \to \beta\gamma, \Gamma \to \beta\Gamma, g \to \beta g, c_n \to \sqrt{\beta} c_n, s_n \to \sqrt{\beta} s_n$, and $t \to t/\beta$, it is clear that lowering $g$ allows the observation of similar phenomena with longer time scale if Q can be further improved.

The second example is given by a one-dimensional electromechanical phonon waveguide fabricated by selective etching of a sacrificial layer through periodically arranged etching holes [31]. Compared to coupled beam resonators, the waveguide structure is advantageous to have larger coupling ($g/\omega_0 > 0.1$) and modulation depth ($\Gamma \sim \Delta\omega_0/\omega_0 \sim 0.015$) [49], which are controllable through the device design. The demonstration of the topological soliton discussed in this paper is feasible using such an existing semiconductor-based electromechanical architecture.

It is also interesting to extend this study to generate topological soliton in other systems. For example, the bistablity in a Duffing resonator array can be used to form a kink between two domains of high-amplitude and low-amplitude states when the nonlinear resonators are mutually coupled. Especially, the rotations in phase space between these two fixed points are opposite, and different signs of helicity can be formed. The extension to two-dimensional arrays and the effect of disorder and fluctuation are also interesting subjects, but they require further study and will need to be addressed elsewhere.

VI. Conclusion

Propagation of a topological soliton in a one-dimensional array of coupled parametric resonators was theoretically studied. Both in continuous and discrete models, a kink between two domains with different phase states can propagate stably. The travelling solution has a topological nature, showing a fixed chirality in phase space. The effect of damping on the propagation dynamics was studied, and the velocity of the kink was reduced by the damping. However, even with a finite damping, an additional harmonic excitation lifts the symmetry between two phase states and drives the soliton propagation at fixed velocity. Collision between the kink and antikink solutions is studied and it generates a localized breather mode, while the two solitons disappear if their energies are not large enough. These finding demonstrate the feasibility of using existing semiconductor-based electromechanical resonator array.

ACKNOWLEDGMENTS

We thank Dr. H. Okamoto, Dr. D. Hatanaka, and Dr. M. Kusoru for their support and helpful comments on the feasibility using devices. We are also grateful to Dr. R. Ohta and Dr. M. Asano for fruitful discussions.